\pdfoutput=1

\documentclass[11pt,twoside,a4paper,cmspaper,final,collab]{cms-tdr}
\def\svnVersion{436780P}\def\cmsCernNoTag{CERN-EP-2016-300}\def\cmsCernDate{\today}\def\cmsMessage{Published in Physics Letters B as \href{http://dx.doi.org/10.1016/j.physletb.2017.06.062}{\doi{10.1016/j.physletb.2017.06.062.}}}

\begin{document}\cmsNoteHeader{EXO-16-025}

\hyphenation{had-ron-i-za-tion}
\hyphenation{cal-or-i-me-ter}
\hyphenation{de-vices}
\RCS$Revision: 399331 $
\RCS$HeadURL: svn+ssh://svn.cern.ch/reps/tdr2/papers/EXO-16-025/trunk/EXO-16-025.tex $
\RCS$Id: EXO-16-025.tex 399331 2017-04-14 20:05:39Z alverson $
\newlength\cmsFigWidth
\ifthenelse{\boolean{cms@external}}{\setlength\cmsFigWidth{0.85\columnwidth}}{\setlength\cmsFigWidth{0.4\textwidth}}
\ifthenelse{\boolean{cms@external}}{\providecommand{\cmsLeft}{top\xspace}}{\providecommand{\cmsLeft}{left\xspace}}
\ifthenelse{\boolean{cms@external}}{\providecommand{\cmsRight}{bottom\xspace}}{\providecommand{\cmsRight}{right\xspace}}
\ifthenelse{\boolean{cms@external}}{\providecommand{\cmsRRight}{Bottom\xspace}}{\providecommand{\cmsRRight}{Right\xspace}}
\ifthenelse{\boolean{cms@external}}{\providecommand{\cmsLLeft}{Top\xspace}}{\providecommand{\cmsLLeft}{Left\xspace}}
\newcommand{\zg}{\ensuremath{\Z\gamma}\xspace}
\newcommand{\xzg}{\ensuremath{\mathrm{X}\to\Z\gamma}\xspace}
\newcommand{\MJ}{\ensuremath{M_\mathrm{j}}\xspace}
\newcommand{\MZg}{\ensuremath{M_{\Z\gamma}}\xspace}
\providecommand{\MGAMCNLO} {\MADGRAPH{}5\_a\MCATNLO\xspace}

\cmsNoteHeader{EXO-16-025}
\title{\texorpdfstring{Search for high-mass $\Z\gamma$ resonances in proton-proton collisions at  $\sqrt{s}=8$ and 13\TeV using jet substructure techniques}{Search for high-mass Z gamma resonances at sqrt(s) = 8 and 13 TeV using jet substructure techniques}}

\date{\today}

\abstract{
A search for massive resonances decaying to a Z boson and a photon is performed
in events with a hadronically decaying Z boson candidate, separately in light-quark and b quark decay modes, identified using jet substructure and advanced b tagging techniques.
Results are based on samples of proton-proton collisions collected with the CMS detector at the LHC at center-of-mass energies of 8 and $13\TeV$,
corresponding to integrated luminosities of 19.7 and 2.7\fbinv, respectively. The results of the search
are combined with those of a similar search in the leptonic decay modes of the Z boson, based on the same data sets.
Spin-0 resonances with various widths and with masses in a range between $0.2$ and $3.0\TeV$ are considered. No significant excess is observed either in the individual analyses or the combination.
The results are presented in terms of upper limits on the production cross section of such resonances and constitute the
most stringent limits to date for a wide range of masses.}

\hypersetup{%
pdfauthor={CMS Collaboration},%
pdftitle={Search for high-mass Z gamma resonances at sqrt(s) = 8 and 13 TeV using jet substructure techniques},%
pdfsubject={CMS},%
pdfkeywords={CMS, physics, resonances, EXO}}

\maketitle

\section{Introduction}
\label{sec:intro}
Searches for resonant production of new particles, postulated in theories beyond the standard model (SM), are a cornerstone of the CERN LHC
physics program. Of particular interest are searches for resonances decaying into a pair of massive SM gauge bosons ($\PW\PW$, $\PW\PZ$, $\PZ\PZ$, with the most recent results described in Refs.~\cite{diboson1,diboson2,diboson3,diboson4}), as well as final states with photons, such as $\PW\gamma$, $\PZ\gamma$, and $\gamma\gamma$. The search in the diphoton final state (together with the results from the $\PW\PW$ and $\PZ\PZ$ channels) played a key role in the discovery of the H(125) boson  by the ATLAS and CMS
Collaborations~\cite{Higgs1,Higgs2,Higgs3} in 2012.

In general, any resonance decaying into the $\gamma\gamma$ or $\PZ\PZ$ channels should also have a $\PZ\gamma$ decay mode, with the relative branching fractions fixed by the SU(2)$_\mathrm{L}$ couplings of the new resonance. Resonances with a spin of 0, 1, or 2 that can decay via the $\PZ\gamma$ channel feature in a variety of proposed theoretical extensions of the SM. Examples include: technicolor~\cite{EL}, extended Higgs boson sector~\cite{BT,LLS}, extra spatial dimension~\cite{DHR,ASS}, and little Higgs~\cite{FS} models, as well other beyond the SM theories. The $\PZ\gamma$ mode is also an important, and yet to be established, decay of the Higgs boson. In particular, the H(125) boson is expected to decay in the $\PZ\gamma$ channel with a branching fraction of 0.16\%, compared to the 0.23\% and 2.67\% branching fractions in the $\gamma\gamma$ and $\PZ\PZ$ channels, respectively~\cite{YR3}. Thus, if a new resonance is observed in one or both of these final states, the analysis of the $\PZ\gamma$ channel may provide crucial information on its nature.

In this Letter we describe a search for spin-0 $\PZ\gamma$ resonances in the hadronic decay channel of the \PZ boson, as well as a combination with the previously published results of an analogous search in the leptonic decay channels~\cite{Zgamma-lep}. The analysis and the combination are based on data sets recorded with the CMS detector at the LHC in proton-proton collisions at center-of-mass energies of 8 and 13\TeV, corresponding to integrated luminosities of 19.7 and 2.7\fbinv, respectively.

We look for a resonance with a relatively narrow width appearing
on top of a smooth \zg invariant mass spectrum constructed with an energetic photon
and with the $\PZ$ boson decay products corresponding to the largest branching fraction: $\PZ\to\qqbar$.
While a search in the leptonic $\PZ$ boson decay modes
has lower SM background, resulting in a higher sensitivity for new resonance masses
less than about $1\TeV$, for higher mass values it is the hadronic $\PZ$ boson decay
channel that dominates the sensitivity.

Depending on the mass of a new resonance, the $\PZ$ boson decay products may be
reconstructed as two resolved jets, or as a single jet resulting from the merging of the two quark jets.
The fraction of events corresponding to the merged topology, which has low SM backgrounds, increases with the mass of the resonance.
 In this analysis we focus on relatively high invariant masses of a hypothetical resonance $\mathrm{X} \to \PZ\gamma$, and therefore consider only the merged jet topology.
We use jet substructure techniques and advanced tagging methods to infer the presence of a subjet originating from b quark fragmentation. This allows us to distinguish a signal from the dominant background from direct photon and QCD multijet production, with one of the jets spuriously passing jet substructure requirements. The background is determined directly from a fit to data.

Previous searches for resonances decaying into $\PZ\gamma$ have been pursued by the L3 Collaboration at the CERN LEP~\cite{L3} and the D0 Collaboration at the Fermilab Tevatron~\cite{D0-1,D0-2}. At the LHC, searches for such resonances have been carried out by the ATLAS Collaboration at $\sqrt{s} = 7\TeV$~\cite{ATLAS-Vgamma1} and $8\TeV$~\cite{ATLAS-Vgamma2} in the context of technicolor-like spin-1 resonances or extended Higgs sector spin-0 resonances, as well as by the ATLAS and CMS Collaborations using the combined 7 and $8\TeV$ data sets in the context of a search for $\PH\to \PZ\gamma$ decay~\cite{CMS-HZg,ATLAS-HZg}. All these analyses have been performed in the dilepton ($\Pep\Pem$ and $\PGmp\PGmm$) decay channels of the \PZ boson. Recently, the ATLAS Collaboration completed a search for high-mass spin-0 $\PZ\gamma$ resonances at $\sqrt{s} = 13\TeV$ in the combination of leptonic and hadronic \PZ boson decay channels, also using jet substructure techniques, but without identification of b quarks within the jet~\cite{ATLAS-Vgamma3}.
\section{The CMS detector}
\label{sec:det}
The central feature of the CMS apparatus is a superconducting solenoid of 6\unit{m} internal diameter, providing a magnetic field of 3.8\unit{T}. Within the solenoid volume are a silicon pixel and strip tracker, a lead tungstate crystal electromagnetic calorimeter (ECAL), and a brass and scintillator hadron calorimeter (HCAL), each composed of a barrel and two endcap sections. Forward calorimeters extend the pseudorapidity coverage provided by the barrel and endcap detectors up to $\abs{\eta} < 5$. Muons are measured in gas-ionization detectors embedded in the steel flux-return yoke outside the solenoid.

The silicon tracker measures charged particles within the pseudorapidity range $\abs{\eta}< 2.5$. For nonisolated particles of $1 < \pt < 10\GeV$ and $\abs{\eta} < 1.4$, the track resolutions are typically 1.5\% in \pt and 25--90 (45--150)\mum in the transverse (longitudinal) impact parameter \cite{TRK-11-001}.

In the region $\abs{ \eta } < 1.74$, the HCAL cells have widths of 0.087 in pseudorapidity and 0.087 in azimuth ($\phi$). In the $\eta$-$\phi$ plane, and for $\abs{\eta} < 1.48$, the HCAL cells map on to $5{\times}5$ arrays of ECAL crystals to form calorimeter towers projecting radially outwards from close to the nominal interaction point. For $\abs{\eta} > 1.74$, the coverage of the towers increases progressively to a maximum of 0.174 in $\Delta \eta$ and $\Delta \phi$.

The first level of the CMS trigger system, composed of custom hardware processors, uses information from the calorimeters and muon detectors to select the most interesting events in a fixed time interval of 3.2\mus. The high-level trigger (HLT) processor farm further decreases the event rate from around 100\unit{kHz} to less than 1\unit{kHz}, before data storage.

A detailed description of the CMS detector, together with a definition of the coordinate system used and the relevant kinematic variables, can be found in Ref.~\cite{Chatrchyan:2008zzk}.

\section{Event selection and Monte Carlo simulation}
\label{sec:eventsel}
The data sets used in the analysis correspond to integrated luminosities of 19.7 and 2.7\fbinv collected with the CMS detector in
pp collisions at the LHC in 2012 ($\sqrt{s} = 8\TeV$) and 2015 ($\sqrt{s} = 13\TeV$), respectively.  Events are selected with an online HLT algorithm, which requires one photon candidate, passing loose identification requirements, with $\pt>150$ (165)\GeV and $\abs{\eta} < 2.5$ in 8 (13)\TeV data.
The trigger is fully efficient for reconstructed photons with $\pt > 170$ (180)\GeV.

In the subsequent analysis, events are reconstructed using a particle-flow (PF) algorithm~\cite{PFT-09-001,PFT-10-001} that identifies each individual particle (photon, electron, muon, charged hadron, and neutral hadron) with an optimized combination of information from the various elements of the CMS detector. The energy of photons is obtained from the ECAL measurement. The energy of electrons is determined from a combination of the electron momentum at the primary interaction vertex as determined by the tracker, the energy of the corresponding ECAL cluster, and the energy sum of all bremsstrahlung photons spatially compatible with originating from the electron track. The energy of muons is obtained from the curvature of the corresponding track. The energy of charged hadrons is determined from a combination of their momentum measured in the tracker and the matching ECAL and HCAL energy deposits, corrected for zero suppression effects and for the response function of the calorimeters to hadronic showers. Finally, the energy of neutral hadrons is obtained from the corresponding corrected ECAL and HCAL energy.

Events are required to have at least one reconstructed collision vertex
within $24\cm$ along the beam axis and $2\cm$
in the plane transverse to the beams of the mean $\Pp\Pp$ interaction position. The vertex with the highest sum of $\pt^2$ of all the associated tracks is taken to be the primary vertex in the event.

Photon candidates are reconstructed from the energy deposits in the ECAL and required to be within the barrel fiducial region of the detector ($\abs{\eta} < 1.4442$) and have $\pt > 170$ (180)\GeV in the 8 (13)\TeV analysis, thus ensuring that the trigger is fully efficient.
Events with a photon reconstructed in the endcap region ($1.566 < \abs{\eta} < 2.5$)
suffer from a large $\gamma$+jet background and do not add to the sensitivity of the analysis;
therefore, they are not considered. Photon identification is based on a multivariate analysis, employing
a boosted decision tree (BDT) algorithm. The input to the BDT algorithm contains
shower shape and isolation variables, as well as variables that account for the dependencies
of the shower shape and isolation variables on the additional interactions in the same or neighboring bunch crossings (pileup)
~\cite{CMS-EGM-14-001}. In addition, a conversion-safe electron veto (CSEV)~\cite{CMS-EGM-14-001} is applied. Isolation variables are computed from PF candidates in a cone of radius $\Delta R =  \sqrt{\smash[b]{(\Delta\phi)^2 + (\Delta\eta)^2}}  < 0.3$ around the photon candidate.
The photon BDT has been trained and optimized separately for 8 and 13\TeV data, so the standard working points are different for the two data sets. In each case we use a working point that corresponds to a typical photon reconstruction and identification efficiency of 90\% in
the photon $\pt$ range used in the analysis.

Large-cone jets are used to reconstruct hadronically decaying Lorentz-boosted $\PZ$ boson candidates in the event.
In both 8 and 13\TeV analyses, they are reconstructed using PF candidates. The 8\TeV analysis employs the Cambridge--Aachen (CA) clustering algorithm~\cite{CA}, while the 13\TeV analysis uses the anti-\kt
algorithm~\cite{Cacciari:2008gp}, both with a distance parameter of 0.8. (The change in the default jet clustering algorithm for 13\TeV data was motivated by commissioning of new  jet substructure triggers, which rely on the faster anti-\kt algorithm.)
Charged hadrons not originating from the primary vertex are not considered in jet clustering. Corrections based on the jet area~\cite{Cacciari:2007fd} are applied to remove the energy contribution of neutral hadrons from pileup interactions. The energy of the jets is further corrected for the response function of the calorimeter. Jet energy corrections are derived from simulation and are confirmed with in situ measurements using the energy balance of dijet and $\gamma$+jet events~\cite{Chatrchyan:2011ds,Khachatryan:2016kdb}. Additional quality criteria are applied to the jets in order to remove
rare spurious noise patterns in the calorimeters, and also to suppress leptons misidentified as jets.
All jets are required to have $\pt > 170$ (200)\GeV and $\abs{\eta} < 2.0$ in the 8 (13)\TeV analysis.
The requirement on the jet $\eta$ suppresses the background from $\gamma$+jet and QCD multijet events and
ensures that the core of the jet is within the tracker volume of the CMS detector ($\abs{\eta} < 2.5$).
The latter requirement is important for subsequent b quark jet tagging.
All jets are required to be separated from the photon candidate in the event by
a minimum distance of $\Delta R > 1.1$.

To identify $\PZ$ boson candidates, the reconstructed jet mass, evaluated after applying a jet pruning
algorithm~\cite{Ellis:2009su,Ellis:2009me}, is used.
The pruning technique reclusters jet constituents and reduces soft, large-angle QCD radiation, which
would increase the mass of the jet. The algorithm first reclusters each jet starting from
its original constituents with the CA algorithm and then discards
soft and wide-angle radiation in each step of the iterative CA procedure. The same pruning algorithm parameters are used for 8 and 13\TeV data~\cite{Khachatryan:2014vla}. The pruned jet mass ($\MJ$) is computed from the sum of the four-momenta of the remaining constituents,
and is corrected with the same factor as the one used to correct the jet energy.
To select a $\PZ$ boson candidate we require the pruned jet mass to be between 70 and 110\GeV (75 and 105\GeV) in 8 (13)\TeV data. We note that the jet mass resolution~\cite{Khachatryan:2014vla} is not sufficient to resolve between the $\PZ$ and $\PW$ bosons decaying hadronically, with the decay products reconstructed as a single large-radius jet. However, since the backgrounds involving $\PW$ bosons are very small, this does not affect the sensitivity of the analysis.

To further discriminate against the direct photon and QCD multijet backgrounds, pruned jets are split into
two subjets by reversing the final iteration in the jet clustering algorithm.
These subjets are classified as those originating or not originating from b quark fragmentation using
the combined secondary vertex (CSV) b tagging algorithm~\cite{BTV,CMS-PAS-BTV-13-001,CMS-PAS-BTV-15-001}.
The jet is identified as one consistent with a $\PZ\to\bbbar$ candidate
if at least one of its subjets satisfies the medium working point of the CSV algorithm and the other subjet
satisfies the loose working point. The medium and loose working points correspond to 70 and 85\% (20 and 50\%) b quark jet tagging efficiency for $\pt < 300\GeV$ ($\pt=1\TeV$), and 1--2\% (10--15\%) light-flavor quark or gluon jet misidentification rate.
The b tagging efficiency in the simulation is corrected to match the one measured
in data~\cite{CMS-PAS-BTV-13-001,CMS-PAS-BTV-15-001}.

In order to further enhance the signal sensitivity, a requirement on the photon $\pt$ with respect to the reconstructed invariant mass of the \PZ candidate and the photon is imposed: $\pt^\gamma/M_{\mathrm{j} \gamma} > 0.34$. This requirement is similar to a selection on the scattering angle of the $\gamma$+jet system, which peaks at higher values for signal than for the background, particularly in case of spin-0 resonances. The value of the cutoff is chosen to maximize the discovery potential for a narrow resonance over the considered mass range. It has 85--90\% selection efficiency for the signal, and about 65\% selection efficiency for the SM background, which is dominated by $\gamma$+jet events, with the prompt photon and a light-flavored jet misidentified as a large radius, massive jet.

The events with a reconstructed photon and a large-cone jet consistent with a Z boson candidate are split into two categories: with or without a $\PZ\to\bbbar$ candidate. These two categories are mutually exclusive and are used simultaneously in the analysis. The summary of the selections is given in Table~\ref{tab:sel}.

\begin{table*}[htpb]
\begin{center}
\topcaption{Summary of event selection.}
\label{tab:sel}
\begin{tabular}{lccc}
\hline
Requirement & 8\TeV & \hfill & 13\TeV \\
\hline
Trigger & $\pt^\gamma > 150$\GeV, $\abs{\eta^\gamma} < 2.5$ & \hfill & $\pt^\gamma > 165$\GeV, $\abs{\eta^\gamma} < 2.5$\\
\hline
Photon & $\pt^\gamma > 170$\GeV, $\abs{\eta^\gamma} < 1.4442$ & \hfill & $\pt^\gamma > 180$\GeV, $\abs{\eta^\gamma} < 1.4442$ \\
Photon BDT & $>$0.133 ($\sim$90\% efficiency) & \hfill & $>$0.374 ($\sim$90\% efficiency)\\
Jet & $\pt^\mathrm{j} > 170$\GeV, $\abs{\eta^\mathrm{j}} < 2.0$ & \hfill & $\pt^\mathrm{j} > 200$\GeV, $\abs{\eta^\mathrm{j} }< 2.0$\\
Pruned jet mass & $70 < M_\mathrm{j} < 110$\GeV & \hfill &  $75 < M_\mathrm{j} < 105$\GeV\\
$\Delta R_{\mathrm{j} \gamma}$ & \multicolumn{3}{c}{$>$1.1}\\
 $\pt^\gamma/M_{\mathrm{j} \gamma}$ & \multicolumn{3}{c}{$>$0.34}\\
\hline
b-tagged category &  \multicolumn{3}{c}{One subjet passing CSV medium; the other CSV loose}\\
Antitagged category & \multicolumn{3}{c}{Failing the above criterion}\\
\hline
\end{tabular}
\end{center}
\end{table*}

A Monte Carlo (MC) simulation, including the effects of pileup, is used to model the signal invariant mass peak and calculate the signal selection efficiency for various mass hypotheses between 0.65 and $3.0 \TeV$ and for two width assumptions for a spin-0 resonance. One width assumption is termed ``narrow", and has the width set to 0.014\% of the particle mass, and the second is referred to as ``broad" with the width set to 5.6\% of the mass. The first choice corresponds to a resonance with
a natural width much less than the detector resolution. The second width value was
chosen for direct comparison with the ATLAS Collaboration analysis~\cite{ATLAS-750} and corresponds to a resonance somewhat broader than the detector resolution. We assume no interference between the signal and the SM $\PZ\gamma$ background.

Signal samples are generated with the leading order (LO) \PYTHIA 8.205 generator~\cite{Sjostrand:2007gs} using the CTEQ6L~\cite{CTEQ6L} (NNPDF3.0~\cite{Ball:2014uwa}) parton distribution functions (PDFs) for the 8 (13)\TeV analysis. In a second step, the \PYTHIA8 program is used to simulate hadronization and parton showering using the tune 4C~\cite{4C} (CUETP8M1~~\cite{Monash,CUETP8M1}) for the 8 (13)\TeV analysis.

In addition, simulated SM background processes are used to optimize the analysis sensitivity. The SM $\PZ\gamma$ and $\PW\gamma$ backgrounds are small, together less than 2\% of the dominant background from light-flavor jets misidentified as massive jets in $\gamma$+jet and QCD multijet events (in the latter case in addition another jet has to be misidentified as a photon), so we did not use SM $\PZ\gamma$ and $\PW\gamma$ samples for this study. However, this contribution to the background is included in the background estimate from data, as detailed in Section~\ref{sec:backgrounds}. In the 8\TeV analysis, the $\gamma$+jet and QCD multijet events are simulated at LO using \PYTHIA 6.126~\cite{pythia6} with tune Z2*~\cite{CUETP8M1,Z2star1}, while the $\PW$+jets and $\PZ$+jets processes are simulated at LO with \MADGRAPH v5.1.3.30~\cite{MADGRAPH5}. In both cases \PYTHIA6 is used to describe fragmentation and hadronization processes. In the 13\TeV analysis, all these samples are simulated at LO with \MGAMCNLO v2.2.2~\cite{aMCNLO} with the fragmentation and hadronization described by \PYTHIA8.
The CMS detector response is modeled with the \GEANTfour package~\cite{Geant}. The effect of pileup is taken into account by superimposing minimum bias events on the hard scattering, with the multiplicity of additional interactions adjusted to that observed in data. The average pileup in the 8 (13)\TeV data sample was 21 (12). Simulated events are processed with the same chain of reconstruction programs as used for collision data.

\section{Search strategy}
\label{sec:strategy}
The search focuses on the mass range from 0.65 to $3\TeV$.
At the lower edge of this mass range about 50\% of $\PZ$ boson decays correspond to the merged jet topology;
for resonance masses above 2\TeV this fraction exceeds 90\%.
In order to profit from both the high acceptance and low background,
two exclusive search categories are formed in
the analysis: a b-tagged category with a large-cone jet required to be consistent
with the $\PZ\to\bbbar$ decay (as described in the previous section),
and an antitagged category with the large-cone jet failing this requirement. While the branching fraction of $\PZ \to \bbbar$ decay is only about 20\% of all hadronic decays, and there is an additional signal loss due to b tagging inefficiency, the background rejection due to b tagging exceeds a factor of 100. Consequently, the sensitivity of the b-tagged category in the low-mass range with large background is significant, leading to a sizable improvement (as large as 50\%) in the signal sensitivity by splitting an inclusive selection into the two categories.

\begin{figure}[htb!]
\centering
\includegraphics[width=0.49\textwidth]{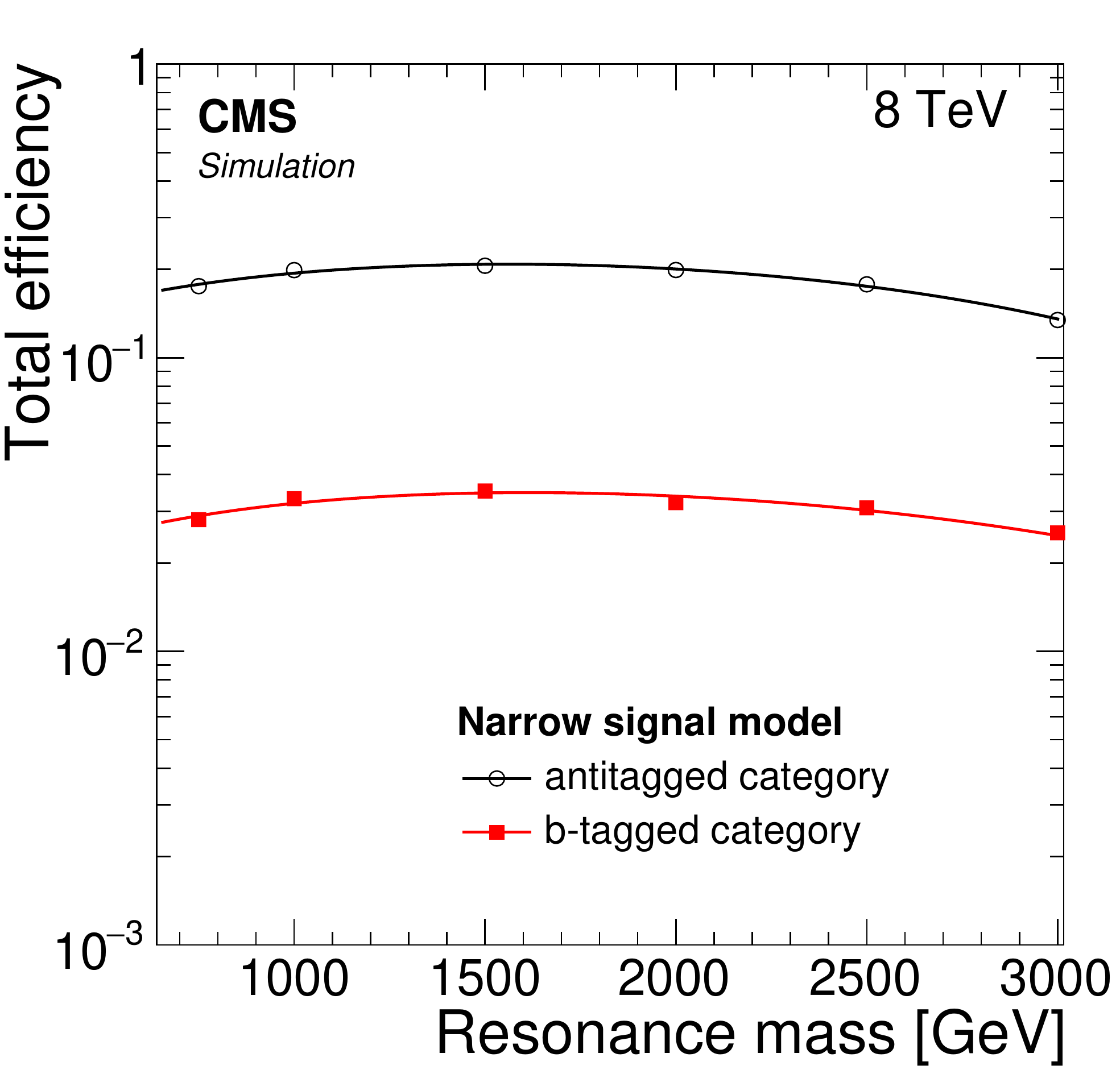}
\includegraphics[width=0.49\textwidth]{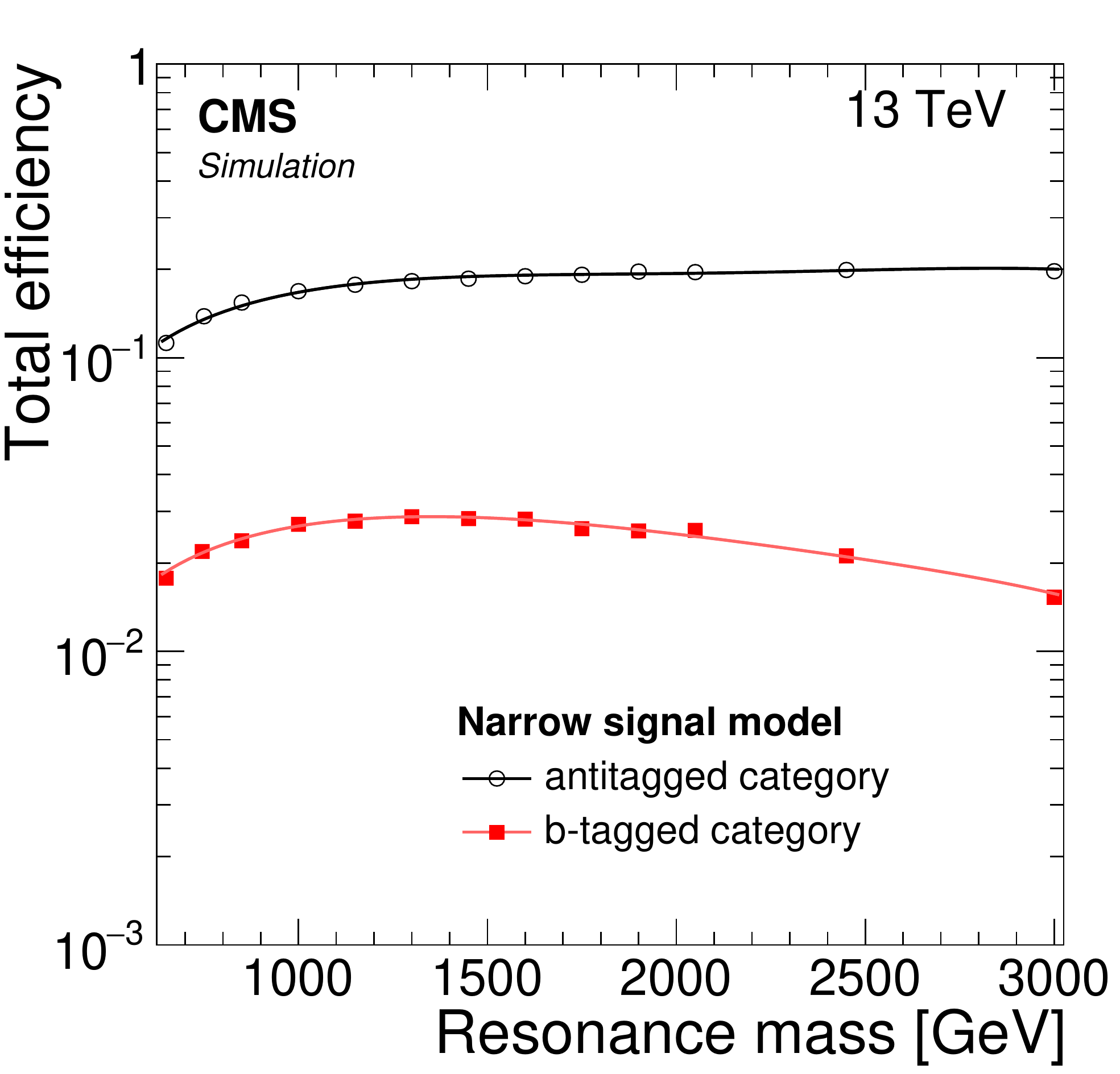}
\caption{Full selection and reconstruction efficiency (including
$\mathcal{B}(\PZ\to\qqbar$)) of the two search categories for a narrow resonance signal
as a function of its mass in the 8\TeV analysis (\cmsLeft) and 13\TeV analysis (\cmsRight).}
\label{fig:signals}
\end{figure}

Figure~\ref{fig:signals} shows the total selection and reconstruction efficiency for the $\xzg$
decay mode of a narrow resonance in the two analysis categories. The total signal efficiency
increases from 17\% (12\%) at 0.65\TeV to 20\% (20\%) at 2\TeV in the antitagged category for the 8 (13)\TeV analysis,
and is between 2 and 3\% for the masses below 2\TeV in the b-tagged category. At very high resonance masses the
b tagging efficiency drops owing to the inability of the b tagging algorithm to disentangle individual jet components
among highly boosted decay products. For a broad resonance at high masses ($>$1.5\TeV) the effect of rapidly falling PDFs introduces a lower tail in the mass distribution. The exact characteristics of this tail are very sensitive to the description of the resonance line-shape. Therefore, in this search, we require that the mass of the resonance corresponds to the core of the distribution, defined as a window centered on the maximum of the Crystal Ball~\cite{CrystalBallRef} (CB) function. The window width is given by $\pm 5$ times the CB function parameter $\sigma$, describing the standard deviation of its Gaussian core. As a result, the efficiency of the analysis requirements, which include this request, for a heavy and broad resonance is lower than for a narrow one and drops to about 3\% (0.3\%) for the antitagged (b-tagged) category for a resonance mass of 3\TeV.

\section{Background and signal modeling}
\label{sec:backgrounds}
Using MC simulation and data studies based on a
lower sideband of the jet mass distribution ($50 < \MJ < 70\GeV$), we observed that the
invariant mass distribution \MZg of the SM background is smoothly
falling and that the distributions of kinematic observables
derived from the lower jet mass sideband match those for the signal selection.
We further checked that the background shapes in the b-tagged and antitagged categories are consistent with each other.

Various families of functions to model the background
shape have been tested in the lower jet mass sideband region, with selection requirements similar to those in the search region. The functions used to fit the background shape in the b-tagged and in the antitagged categories are chosen using the Fisher F-test. This test selects the optimal function by balancing the quality of the fit against the number of parameters required. In each case the following function is chosen:
\begin{equation}
\frac{\rd N}{\rd\MZg} = P_0\,(\MZg/\sqrt{s})^{P_1+P_2\log(\MZg/\sqrt{s})},
\label{eq:Func}
\end{equation}
where $\MZg$ is the invariant mass of the photon and the large-cone jet,
$\sqrt{s}$ is the center-of-mass energy, $P_0$ is a normalization parameter,
and $P_1$, $P_2$ describe the shape of the invariant mass spectrum.

In order to check for the presence of a possible systematic bias from the
choice of the functional form, several tests are carried out with alternative
functional forms, with or without signal injection. For these tests, the mass
spectra in the two analysis categories derived either from the low-mass sideband in data
or from MC simulation are fitted with a variety
of test functions. The shapes obtained in these fits are used to generate
pseudo-datasets with a total number of events randomly drawn from a Poisson
distribution with the mean equal to the yields observed in data. Additionally,
in a set of pseudo-experiments, signals with different mass values and cross sections close to the
expected 95\% confidence level (CL) limits are injected. The full spectrum is fitted with the chosen
function of Eq.~(\ref{eq:Func}) together with a signal model, and the signal
cross section is extracted. Distributions of the difference between the data and the fit divided by the overall uncertainty for the obtained signal
cross sections are constructed, and their shapes are found to be consistent
with a normal distribution with mean less than 0.5 and width
consistent with unity. Thus, we conclude that any possible systematic bias
from the choice of the functional form is small compared to the statistical
uncertainty of the fit, and use the latter
as the only uncertainty in the background prediction.

\begin{figure*}[htpb!]
\centering
\includegraphics[width=0.49\textwidth]{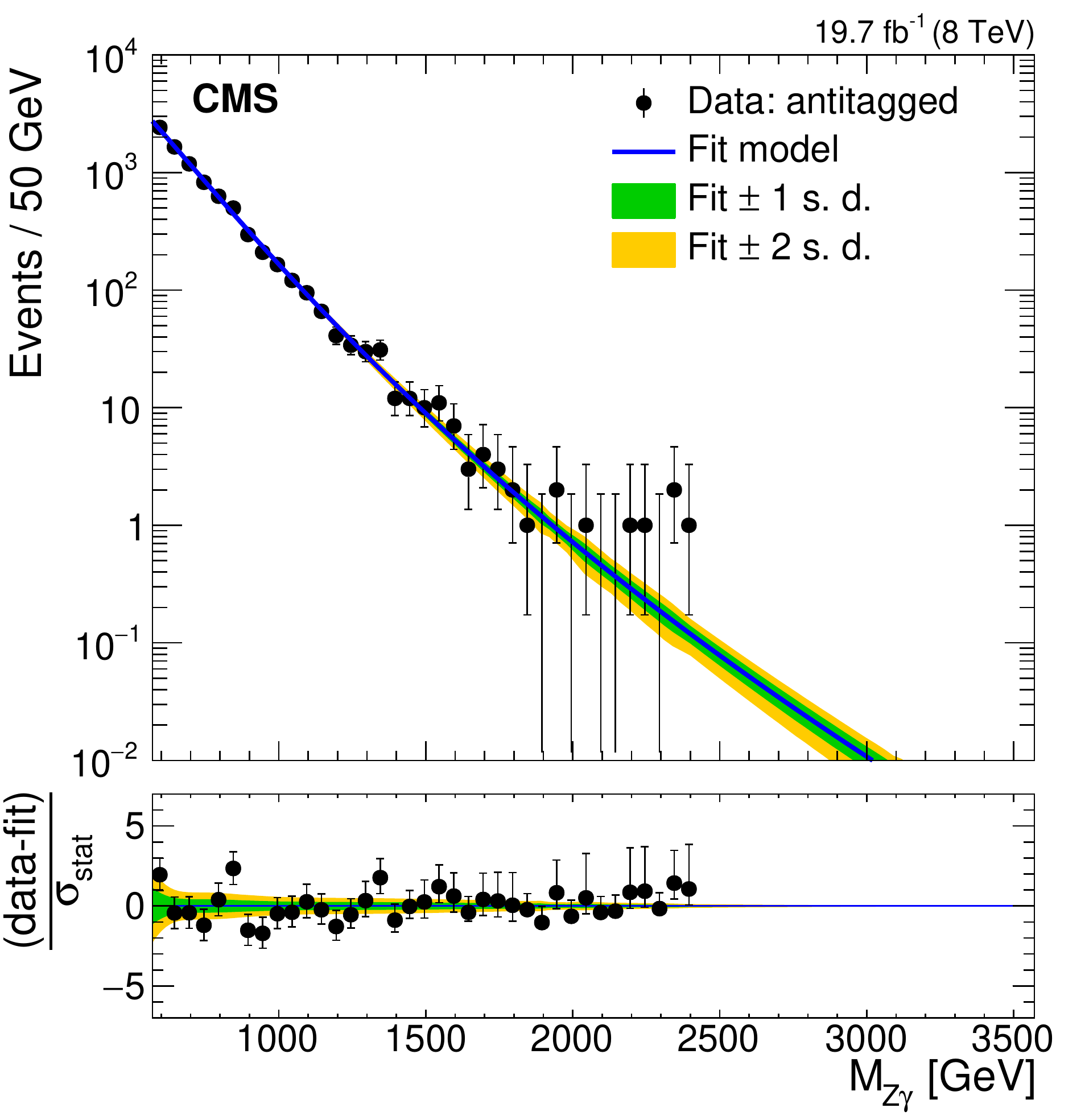}
\includegraphics[width=0.49\textwidth]{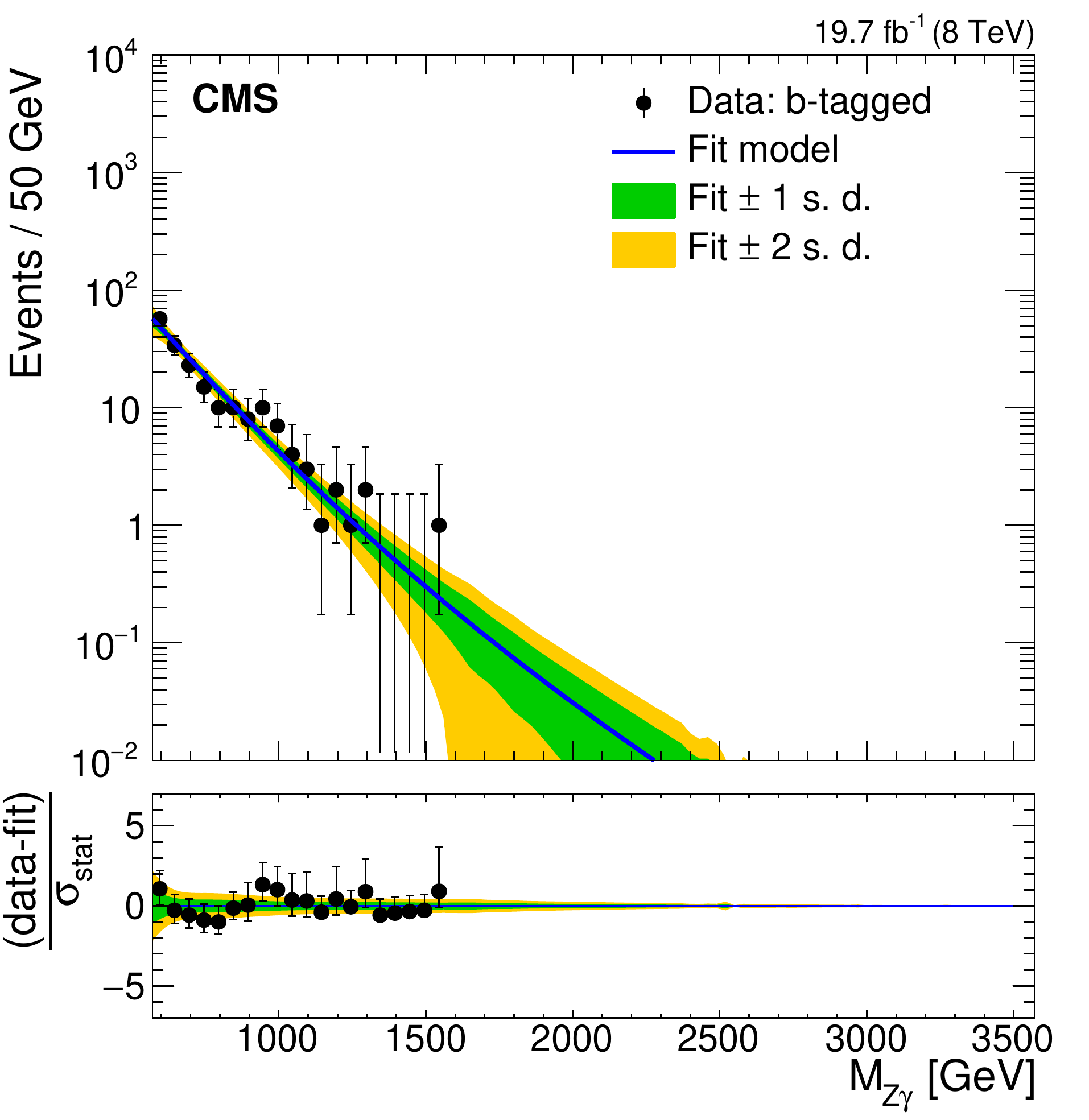}
\includegraphics[width=0.49\textwidth]{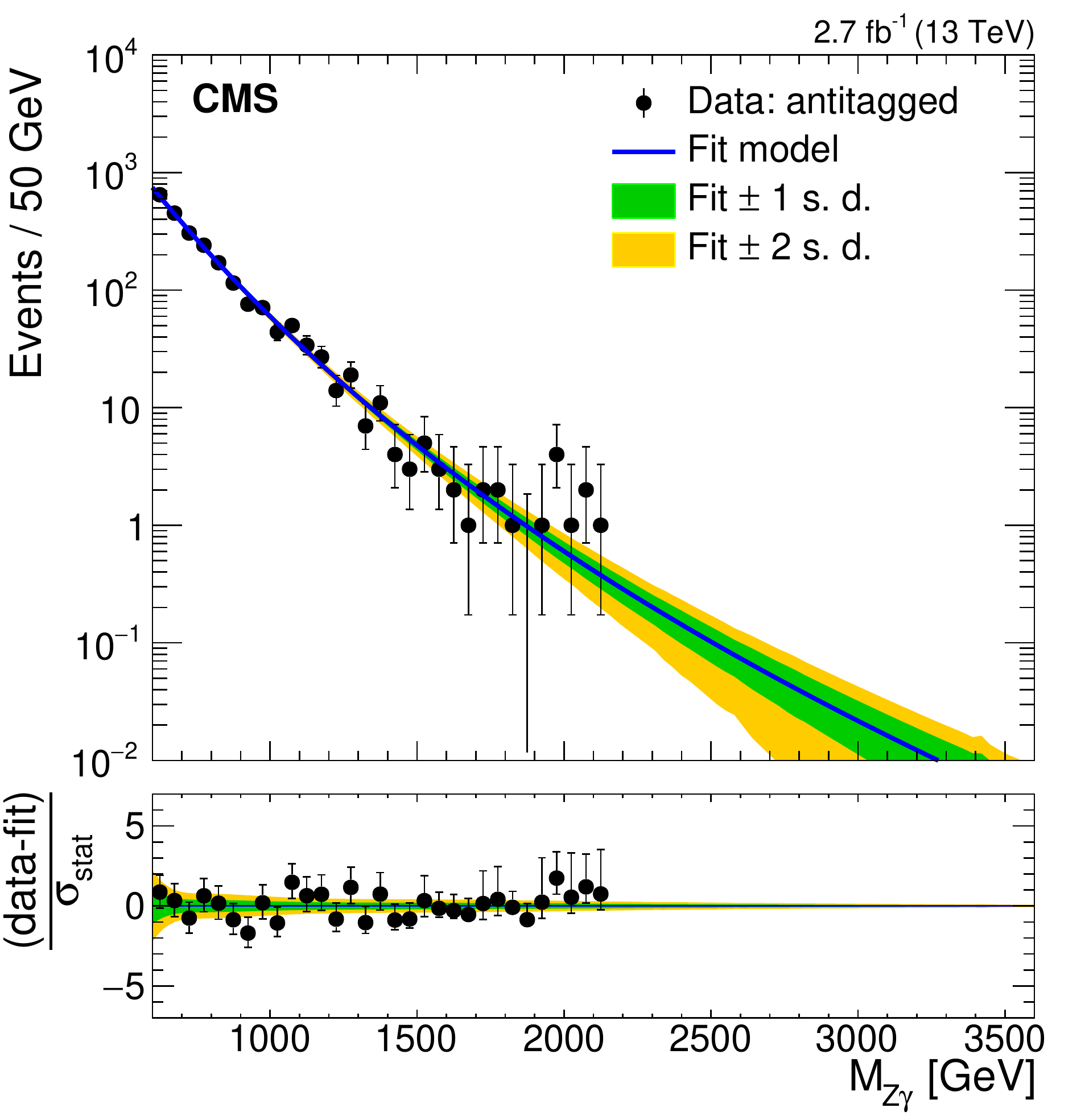}
\includegraphics[width=0.49\textwidth]{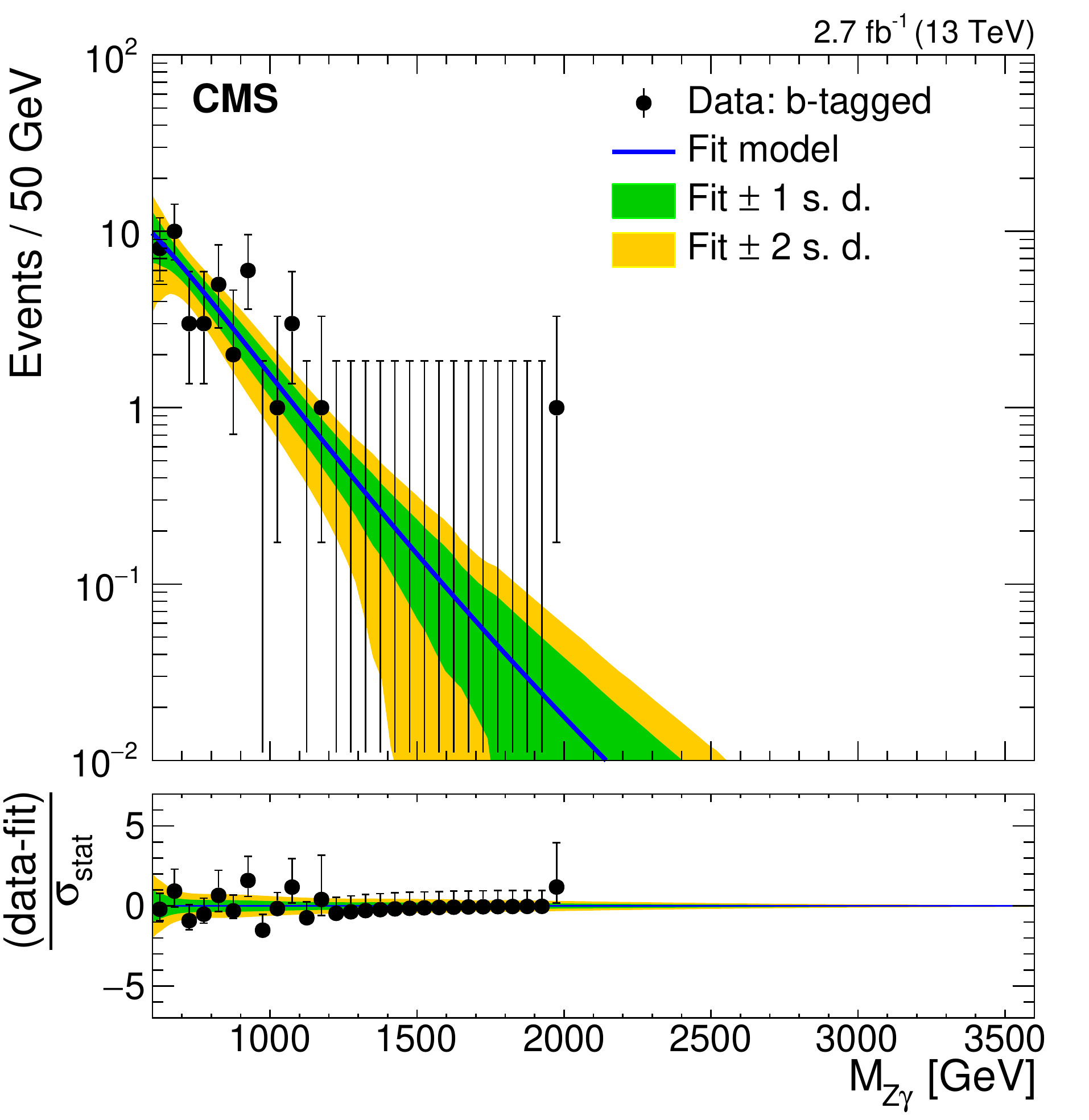}
\caption{Fits to the $\MZg$ invariant mass spectra in the search region for the antitagged (left column) and b-tagged (right column) categories. Upper (lower) row corresponds to 8 (13)\TeV data. The results
of the fits to the two categories with the parametric background shape are shown. The lower panels show the difference between the data and the fit, divided by the statistical uncertainty in data $\sigma_\text{ stat}$. For bins with a low number of data entries, the error bars correspond to the Garwood confidence intervals~\protect\cite{Garwood}. The upper error bars for bins with zero data entries are shown only in the region up to the highest nonzero entry.}
\label{fig:bkgfits}
\end{figure*}

The observed $\MZg$ invariant mass distributions in data in the antitagged and b-tagged categories
along with the corresponding fits are shown
in Fig.~\ref{fig:bkgfits}, separately for 8 and 13\TeV data.

The signal shape is extracted from MC simulation for various
signal hypotheses tested in the analysis. The shape is parameterized with the combination of a
CB function and a Gaussian function in order to ensure a good description of the tails.
To derive the signal shapes for
the intermediate mass values where simulation points are not available, a linear morphing~\cite{morphing} of the shapes obtained from
the MC simulation is used. The typical $\MZg$ resonance core width is found to be 3 and 5\% of the
resonance mass for the narrow and broad resonance hypotheses, respectively.

\section{Systematic uncertainties}
\label{sec:systematics}
Since the background estimation is obtained from a fit to the data,
the only source of the systematic uncertainty in the background estimate
is associated with the possible bias in the choice of the fit function.
This potential bias is checked as described in Section~\ref{sec:backgrounds}
and is found to be negligible with respect to the statistical uncertainty in
the background normalization.

For the signal selection efficiency, there are several sources of
systematic uncertainties, which are summarized in Table~\ref{tab:systSummary}.
Most of the uncertainties affect the overall signal efficiency,
and only the b tagging
efficiency uncertainty can result in signal category migration. The latter is larger
for the 13\TeV analysis owing to the relatively small sizes of control samples in data available for their derivation.

The uncertainties in the jet energy scale and resolution~\cite{Chatrchyan:2011ds,Khachatryan:2016kdb}
are propagated to all relevant quantities, and affect both
the signal yield and its shape. The overall effect of these uncertainties
is found by changing the four-momenta of the jets by an amount equal to the uncertainty in their energy scale, or by smearing them with a resolution function, and
carrying out the full analysis with modified quantities. The
corresponding uncertainties in the signal yield are approximately 2.0 and 2.5\%, respectively.

To account for a slight dependence of the pruned jet mass scale on
the jet $\pt$, an uncertainty in
the Z boson tagging efficiency of 10 (5)\% is applied in the 8 (13)\TeV analysis.

The systematic uncertainties in the photon energy scale and
identification efficiency are derived from $\PZ\to\EE$ events.
The uncertainty in the photon energy scale is found
to be about 1\% and it includes the uncertainty on the extrapolation to
higher-$\pt$ photons.
The ratio between the photon reconstruction and identification efficiencies
in data and in the simulation is consistent with unity within the 2\%
systematic uncertainty
up to a photon $\pt$ of 200\GeV, and within the 4\% systematic uncertainty
in the photon $\pt$ range from $0.2$ to $1.0\TeV$.

The uncertainties in the measurement of the integrated luminosity (2.6\%~\cite{LUM-13-001} and 2.7\% ~\cite{LUM-15-001} in the 8 and 13\TeV analyses, respectively), trigger efficiency (2\%), and pileup description (5\%)
affect the overall signal yield and are taken into account. Concerning the PDF modeling~\cite{Ball:2014uwa}, only the resultant uncertainty in the signal acceptance (2\%), and not the signal cross section, is included in the overall experimental uncertainty.

\begin{table*}[h!]
\centering
\topcaption{Summary of the sources of systematic uncertainties, their magnitudes, effects on the signal yield, and affected quantities. The third column
indicates the magnitude of the yield variation. The last column indicates if the source of the uncertainty
affects the total signal yield, signal shape, or introduces a category migration. The numbers in parentheses correspond to the 13\TeV analysis (whenever there is a difference from the 8\TeV numbers).}
\label{tab:systSummary}
\begin{tabular}{lccc}
\hline
Source          & Magnitude     & Effects on the yield & Affected quantity \\\hline
Jet energy scale           & 2\%          & 2\%     &      yield  \& shape  \\
Jet energy resolution           & 8--10\%          & 3.2 (2.8)\%     &      yield \& shape   \\
\MJ mass range & 10 (5)\%    & 10 (5)\%  &       yield    \\
b tagging efficiency & 5--30\% (10--60\%)    & 4--15\% (15--35\%)  &  migration   \\
\hline
Photon energy scale, resolution & 1\%     & 1\%  & yield \& shape \\
Photon efficiency   & 0.2 (2)\%     & 0.2 (2)\%  & yield \\
Electron veto efficiency   & 0.5 (2.5)\%     & 0.5 (2.5)\%  & yield \\
Photon efficiency extrapolation   & 2\%     & 2\%  & yield \\
Trigger efficiency       & 2\%      & 2\%  & yield \\
\hline
Pileup           & 5\%          & 0.6 (1)\%                    &   yield    \\
Parton distribution functions   & 2\%   & 2\% & yield \\
Integrated luminosity     & 2.7 (2.6)\%       & 2.7 (2.6)\%   &     yield     \\
\hline
\end{tabular}
\end{table*}
\section{Results}
\label{sec:results}
The $\MZg$ invariant mass spectra observed in data in two categories for each data set (8 and 13\TeV), are fitted simultaneously under the background-only, as well as the combined background and signal hypotheses, for various signal mass and width assumptions. Both the 8 and 13\TeV data are well described by the background-only hypothesis. We see no statistically significant evidence for a signal in the entire mass range probed. The largest deviation is seen in 13\TeV data at a mass around 2\TeV with a local significance of 3.6 standard deviations for a narrow resonance hypothesis, which corresponds to a global significance of approximately 2.5 standard deviations assuming a narrow resonance and taking into account the full search range~\cite{LEE}. This excess is not seen in 8\TeV data.
The results are presented as upper limits on the
new resonance production cross section times branching fraction to these
$\PZ\gamma$ final state. The limits are computed at 95\% CL, using the asymptotic approximation~\cite{Gross} of the CL$_\mathrm{s}$
criterion~\cite{Read:2002hq,Junk:1999kv,ATL-PHYS-PUB-2011-011}.
Log-normal prior distributions for  parameters are used to account
for the systematic uncertainties in the signal and background yields, which are
described in Section~\ref{sec:systematics} and summarized in
Table~\ref{tab:systSummary}.

\begin{figure*}[htpb!]
\centering
\includegraphics[width=0.49\textwidth]{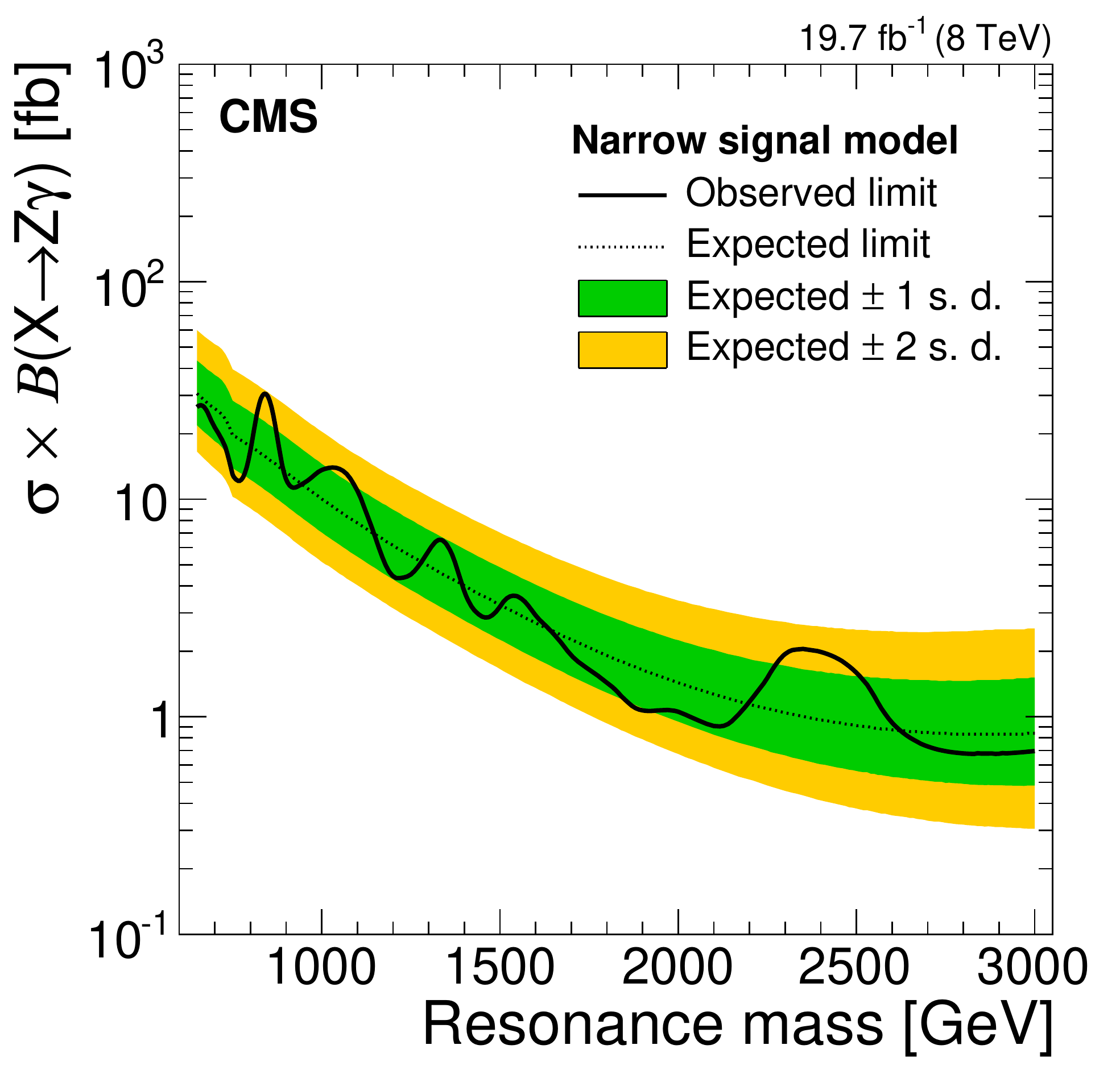}
\includegraphics[width=0.49\textwidth]{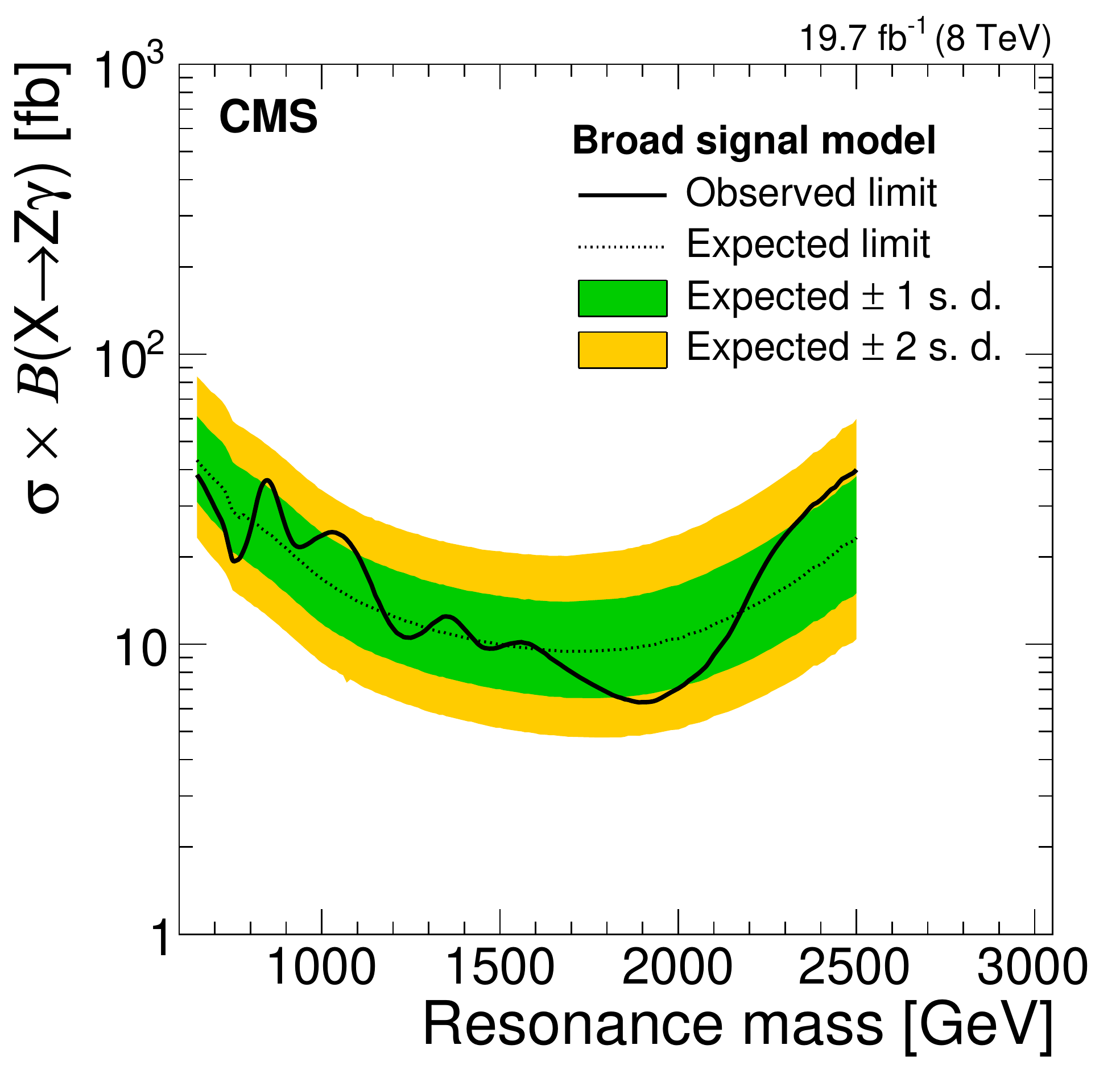}
\includegraphics[width=0.49\textwidth]{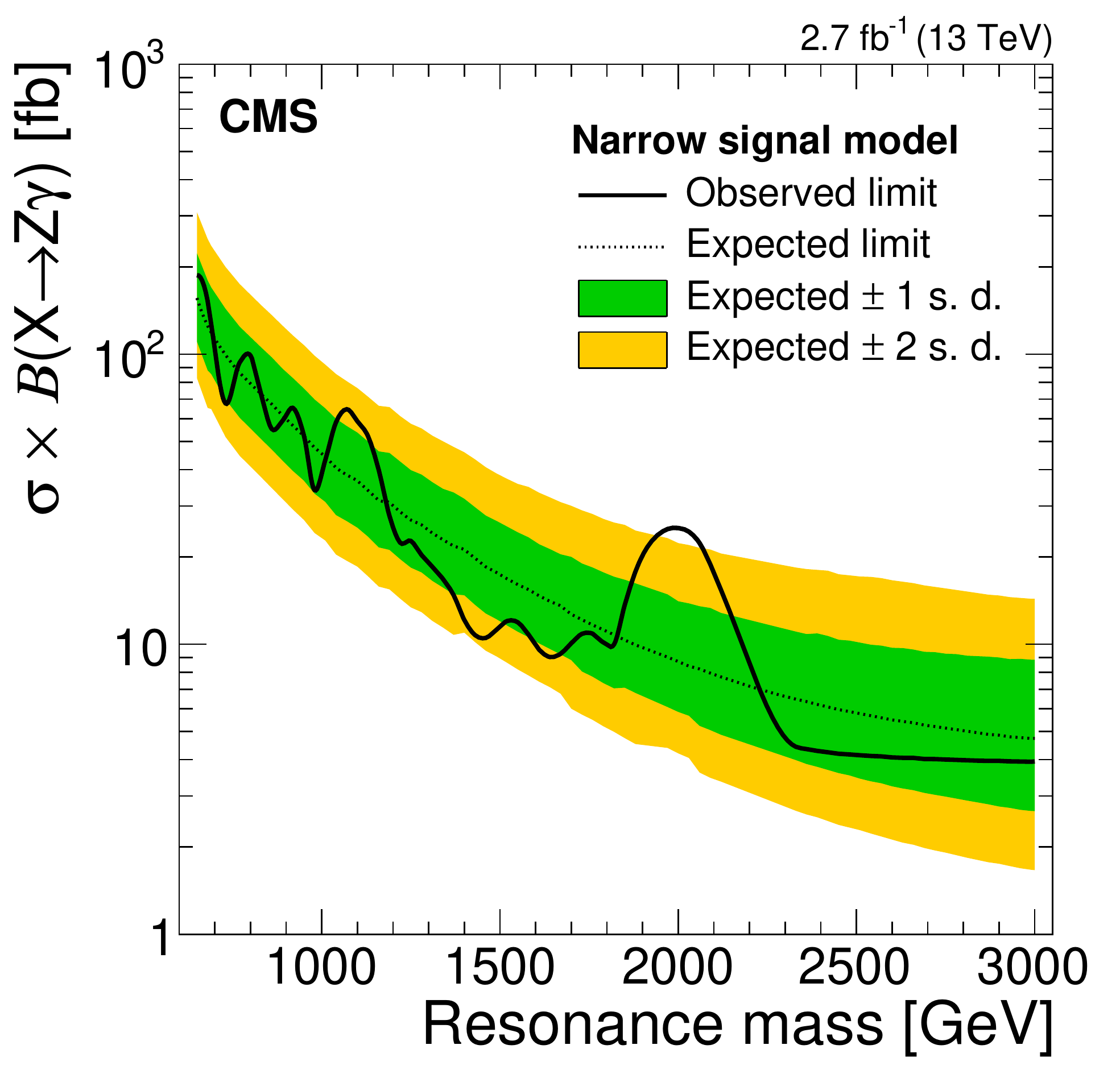}
\includegraphics[width=0.49\textwidth]{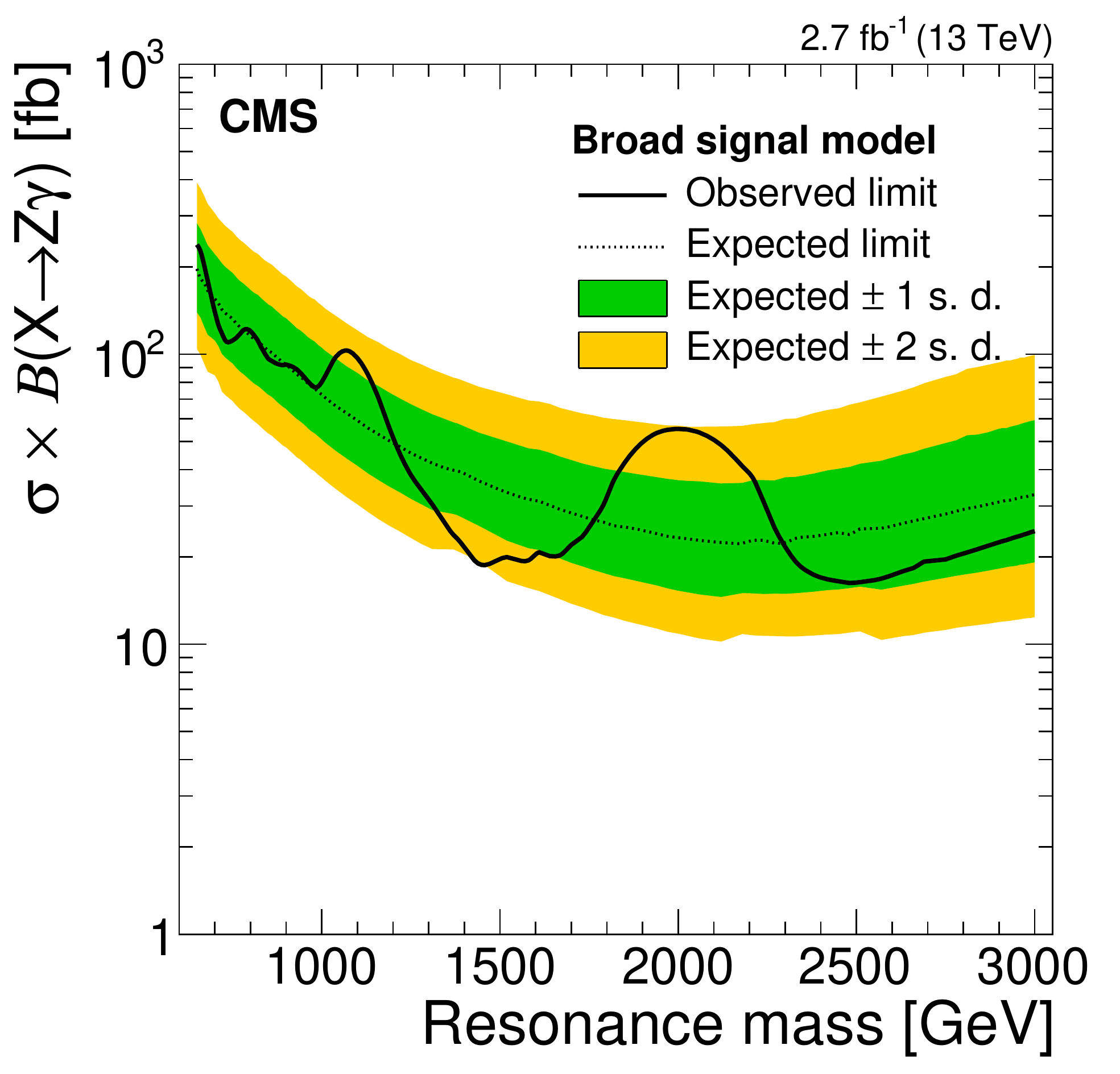}
\caption{Expected and observed upper limits on the product of the cross section and branching fraction ${\cal B}(\mathrm{X} \to \PZ\gamma)$ for the production of a narrow (left) or broad (right) spin-0 resonance, obtained from the combination of antitagged and b-tagged categories in 8\TeV (upper) and 13\TeV (lower) data.}
\label{fig:limits}
\end{figure*}

The expected and observed upper cross section limits for spin-0 resonances with the two benchmark widths from the combination of the antitagged and b-tagged categories are presented in Table~\ref{tab:limits} and Fig.~\ref{fig:limits}. The table shows also individual limits from fits to a single category, illustrating the relative weights of the two categories for various masses.

\begin{table}[h!]
\centering
\topcaption{Observed (expected) limits on the production cross section times branching fraction ${\cal B}(\mathrm{X} \to \PZ\gamma)$ for narrow  resonances from each of the two categories of the analysis, as well as from their combination.}
\label{tab:limits}
\begin{tabular}{cccc}\hline
Mass & \multicolumn{3}{c}{Limits [fb]}\\
\cline{2-4}
[\GeVns{}] & Antitagged & b-tagged & Combined \\\hline
\multicolumn{4}{c}{8\TeV analysis}\\
\hline
750   & 23 (28) & 21 (32) & 13 (20) \\
1000  & 9.7 (13) & 36 (17) &  13 (10) \\
2000  & 1.4 (1.7)  & 3.6 (4.7)  & 1.1 (1.5) \\
2500  & 2.0 (1.1)  & 3.5 (3.9)  & 1.6 (0.9) \\
3000  & 0.8 (1.0)  & 4.0 (4.3)  & 0.7 (0.8) \\
\hline
\multicolumn{4}{c}{13\TeV analysis}\\
\hline
750 & 149 (140) & 94 (150) & 74 (93) \\
1000 & 57 (61) & 71 (87) & 38 (46) \\
2000 & 27 (10) & 63 (37) & 25 (8.7) \\
2500 & 4.7 (6.6) & 42 (41) & 4.1 (5.8) \\
3000 & 4.4 (5.2) & 51 (51) & 3.9 (4.7)\\
\hline
\end{tabular}
\end{table}

The 8 and 13\TeV results can be combined assuming a particular production mechanism for a resonance decaying into the $\PZ\gamma$ channel. Similar to the combination of $\PZ\gamma$ searches in the leptonic decay channel of the Z boson~\cite{Zgamma-lep}, we assume that the hypothetical spin-0 resonance is produced exclusively via gluon fusion, which is a natural production mechanism for a spin-0 particle with Yukawa-like couplings to quarks. The combination takes into account the ratio of gluon-gluon parton luminosities at the two center-of-mass energies, as calculated with the NNPDF2.3 PDFs. This ratio increases from approximately 4.1 for an invariant mass of 650\GeV to 23.7 for a mass of 2.5\TeV. It was checked that the uncertainty in the ratio of parton luminosities at 13 and 8\TeV coming from the PDF uncertainties has negligible effect on the combined results in the range of masses probed. The combination is performed with the same CL$_{\rm s}$ criterion as used to obtain results in the individual channels. We assume that all sources of systematic uncertainty, except for the one related to the photon energy scale, are completely uncorrelated between the analyses at the two energies. This is a reasonable assumption, given that the dominant sources of the systematic uncertainty are the statistical uncertainty in the background fit and b tagging efficiency uncertainties in the signal yield, both of which are determined independently in the 8 and 13\TeV data, and therefore are uncorrelated.

\begin{figure}[htpb!]
\centering
\includegraphics[width=0.5\textwidth]{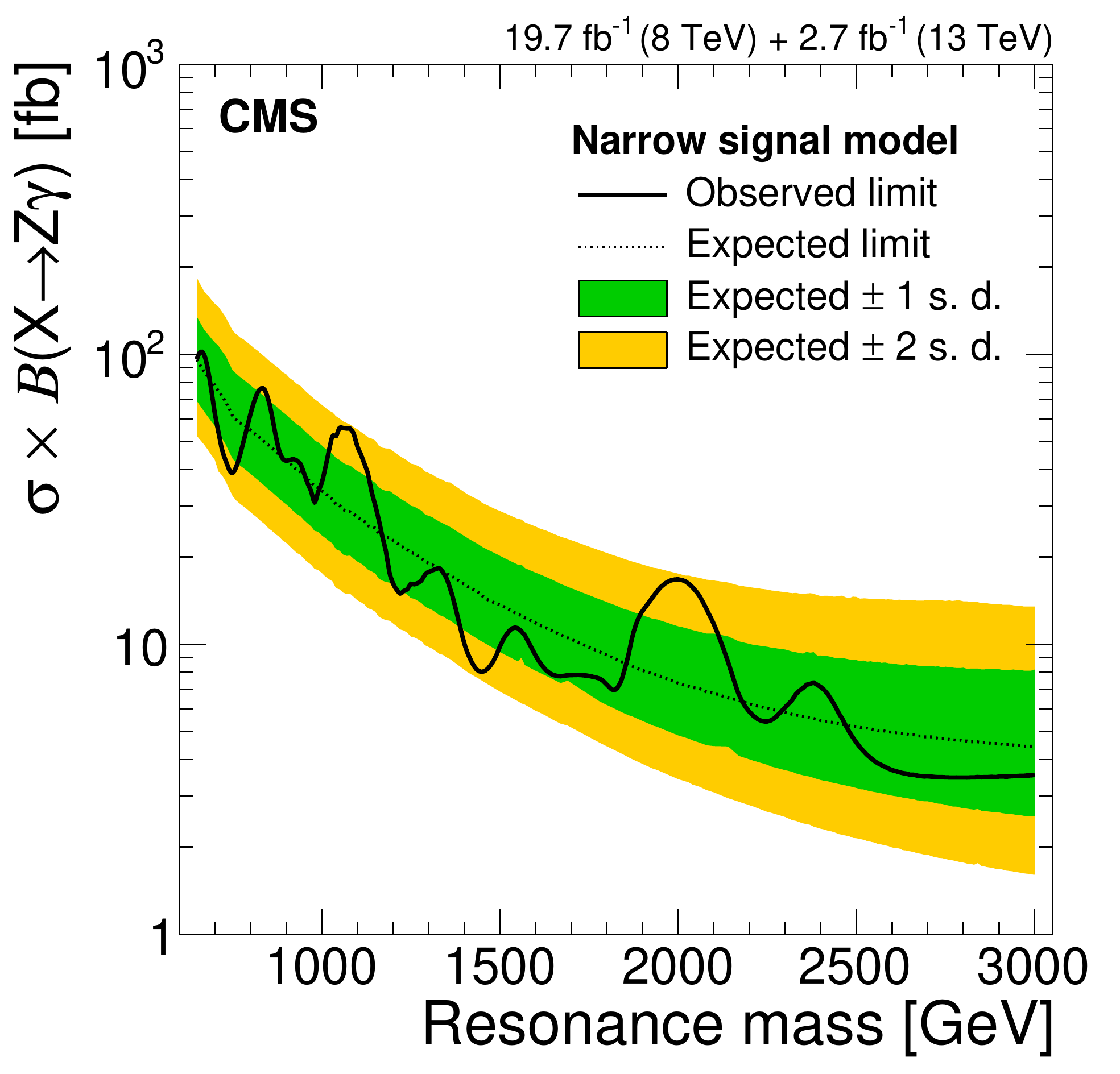}
\caption{Expected and observed limits on the product of the cross section at $\sqrt{s} =13\TeV$ and branching fraction $\mathcal{B}(\mathrm{X} \to \PZ\gamma)$ for the production of a narrow spin-0 resonance, obtained from the combination of the 8 and 13\TeV analyses in the hadronic decay channel, assuming a gluon fusion production mechanism.
}
\label{fig:limitshad}
\end{figure}

The results are expressed in terms of upper limits on a new resonance production cross section via gluon-gluon fusion mechanism at a center-of-mass energy of 13\TeV times branching fraction of the resonance decay in the $\PZ\gamma$ channel. The combined expected and observed 95\% CL limits for narrow resonance production are shown in Fig.~\ref{fig:limitshad}.

\begin{figure}[htb!]
\centering
\includegraphics[width=0.495\textwidth]{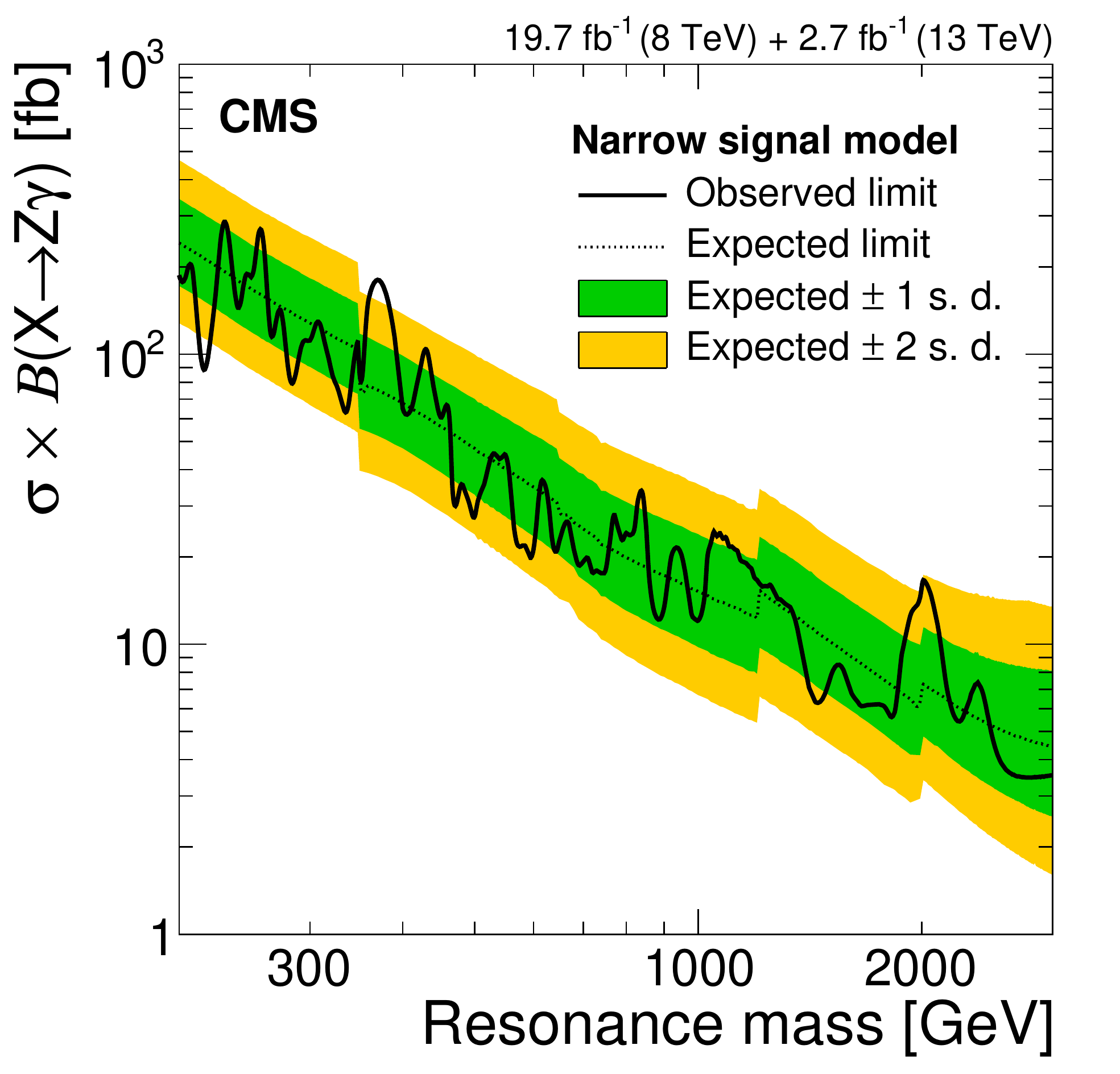}
\includegraphics[width=0.495\textwidth]{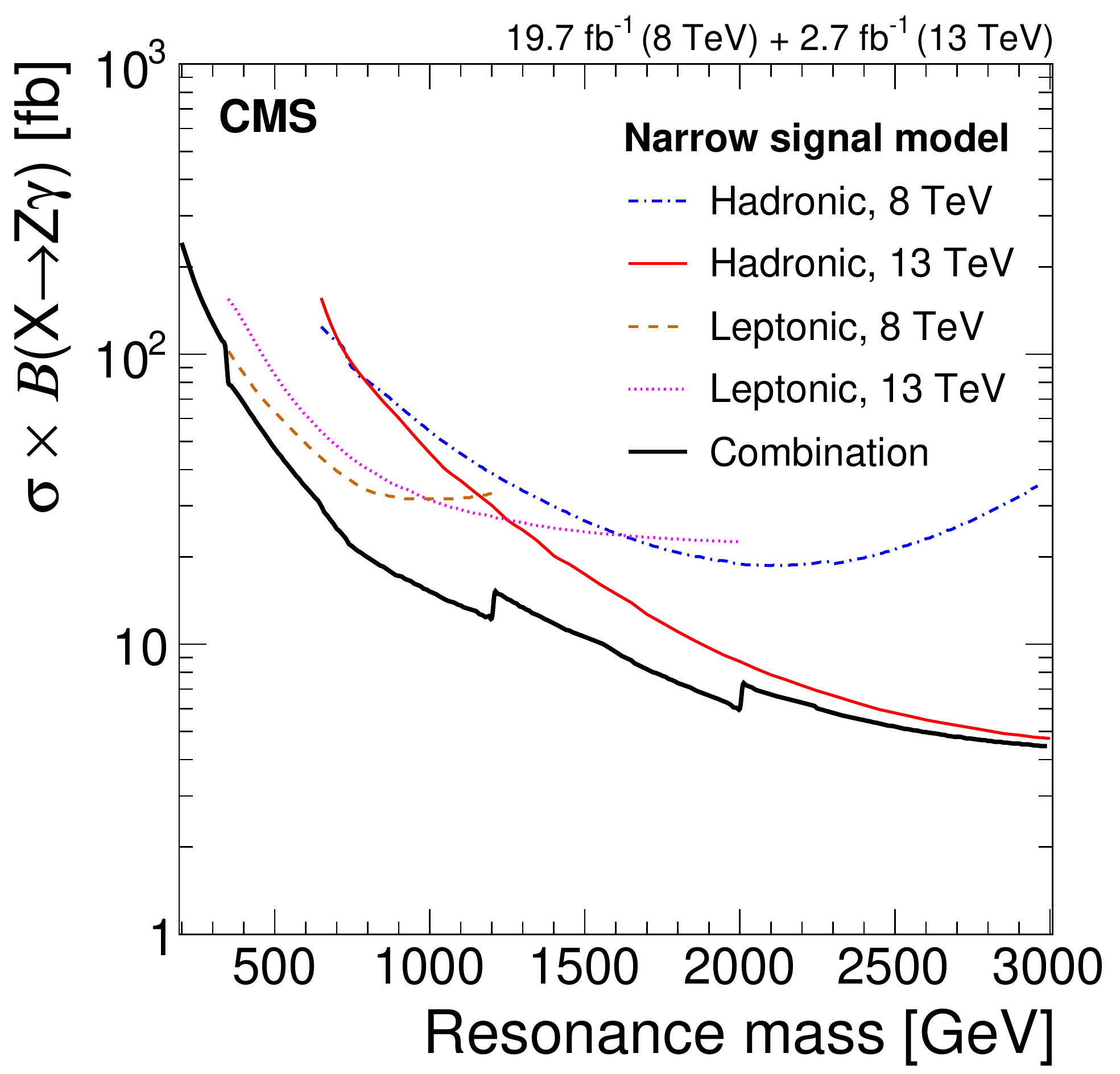}
\caption{\cmsLLeft: Expected and observed limits on the product of the cross section at $\sqrt{s} =13\TeV$ and branching fraction $\mathcal{B}(\mathrm{X} \to \PZ\gamma)$ for the production of a narrow spin-0 resonance, obtained from the combination of the 8 and 13\TeV analyses in hadronic and leptonic~\protect\cite{Zgamma-lep} decay channels of the Z boson, assuming a gluon fusion production mechanism. \cmsRRight: expected limits from the individual and combined analyses, showing the relative contribution of each channel. The discontinuities are due to the difference in the mass ranges used in the individual searches.}
\label{fig:limitscomb}
\end{figure}

The results can be further combined with those from an analogous combined analysis in the leptonic $\PZ$ boson decay channels~\cite{Zgamma-lep}, using the same technique and assumptions. The results are shown in Fig.~\ref{fig:limitscomb}, assuming uncorrelated uncertainties between the leptonic and hadronic channels, except for the uncertainties in the integrated luminosity, PDFs, and photon energy scale and resolution, which are taken as fully correlated between the two analyses. Since the leptonic analysis uses a different photon identification algorithm, the photon efficiency uncertainties are expected to be uncorrelated between the leptonic and hadronic channels.

\section{Summary}
\label{sec:summary}
We have presented a search for new spin-0 resonances decaying to a
\PZ boson and a photon, where the \PZ boson decays hadronically, in the mass range
from $0.65$ to $3.0\TeV$, using 2012 and 2015 proton-proton collision data at center-of-mass energies of 8 and 13\TeV, respectively.
The search is carried out with two exclusive categories of events, with or
without identification of the $\PZ\to\bbbar$ decay, and the final result is
obtained from the combination of these two categories. Jet substructure
and subjet b tagging techniques are used in order to enhance the sensitivity of the analysis. No significant deviation from the standard model prediction is found. Results are presented as upper
limits at 95\% confidence level on the product of the production cross section and the branching fraction of the $\PZ\gamma$ decay channel of a new resonance.
The results of the searches at the two center-of-mass energies are combined assuming the mechanism for production of a new resonance is gluon fusion. These results are further combined with those of analogous searches in the leptonic decay channel of the \PZ boson. The limits set in this analysis are the most stringent limits to date on $\PZ\gamma$ resonances in a wide range of masses.

\begin{acknowledgments}
\label{sec:ack}
We congratulate our colleagues in the CERN accelerator departments for the excellent performance of the LHC and thank the technical and administrative staffs at CERN and at other CMS institutes for their contributions to the success of the CMS effort. In addition, we gratefully acknowledge the computing centres and personnel of the Worldwide LHC Computing Grid for delivering so effectively the computing infrastructure essential to our analyses. Finally, we acknowledge the enduring support for the construction and operation of the LHC and the CMS detector provided by the following funding agencies: BMWFW and FWF (Austria); FNRS and FWO (Belgium); CNPq, CAPES, FAPERJ, and FAPESP (Brazil); MES (Bulgaria); CERN; CAS, MoST, and NSFC (China); COLCIENCIAS (Colombia); MSES and CSF (Croatia); RPF (Cyprus); SENESCYT (Ecuador); MoER, ERC IUT and ERDF (Estonia); Academy of Finland, MEC, and HIP (Finland); CEA and CNRS/IN2P3 (France); BMBF, DFG, and HGF (Germany); GSRT (Greece); OTKA and NIH (Hungary); DAE and DST (India); IPM (Iran); SFI (Ireland); INFN (Italy); MSIP and NRF (Republic of Korea); LAS (Lithuania); MOE and UM (Malaysia); BUAP, CINVESTAV, CONACYT, LNS, SEP, and UASLP-FAI (Mexico); MBIE (New Zealand); PAEC (Pakistan); MSHE and NSC (Poland); FCT (Portugal); JINR (Dubna); MON, RosAtom, RAS and RFBR (Russia); MESTD (Serbia); SEIDI and CPAN (Spain); Swiss Funding Agencies (Switzerland); MST (Taipei); ThEPCenter, IPST, STAR and NSTDA (Thailand); TUBITAK and TAEK (Turkey); NASU and SFFR (Ukraine); STFC (United Kingdom); DOE and NSF (USA).

\hyphenation{Rachada-pisek} Individuals have received support from the Marie-Curie programme and the European Research Council and EPLANET (European Union); the Leventis Foundation; the A. P. Sloan Foundation; the Alexander von Humboldt Foundation; the Belgian Federal Science Policy Office; the Fonds pour la Formation \`a la Recherche dans l'Industrie et dans l'Agriculture (FRIA-Belgium); the Agentschap voor Innovatie door Wetenschap en Technologie (IWT-Belgium); the Ministry of Education, Youth and Sports (MEYS) of the Czech Republic; the Council of Science and Industrial Research, India; the HOMING PLUS programme of the Foundation for Polish Science, cofinanced from European Union, Regional Development Fund, the Mobility Plus programme of the Ministry of Science and Higher Education, the National Science Center (Poland), contracts Harmonia 2014/14/M/ST2/00428, Opus 2013/11/B/ST2/04202, 2014/13/B/ST2/02543 and 2014/15/B/ST2/03998, Sonata-bis 2012/07/E/ST2/01406; the Thalis and Aristeia programmes cofinanced by EU-ESF and the Greek NSRF; the National Priorities Research Program by Qatar National Research Fund; the Programa Clar\'in-COFUND del Principado de Asturias; the Rachadapisek Sompot Fund for Postdoctoral Fellowship, Chulalongkorn University and the Chulalongkorn Academic into Its 2nd Century Project Advancement Project (Thailand); and the Welch Foundation, contract C-1845.
\end{acknowledgments}
\bibliography{auto_generated}

\cleardoublepage \appendix\section{The CMS Collaboration \label{app:collab}}\begin{sloppypar}\hyphenpenalty=5000\widowpenalty=500\clubpenalty=5000\textbf{Yerevan Physics Institute,  Yerevan,  Armenia}\\*[0pt]
A.M.~Sirunyan, A.~Tumasyan
\vskip\cmsinstskip
\textbf{Institut f\"{u}r Hochenergiephysik,  Wien,  Austria}\\*[0pt]
W.~Adam, E.~Asilar, T.~Bergauer, J.~Brandstetter, E.~Brondolin, M.~Dragicevic, J.~Er\"{o}, M.~Flechl, M.~Friedl, R.~Fr\"{u}hwirth\cmsAuthorMark{1}, V.M.~Ghete, C.~Hartl, N.~H\"{o}rmann, J.~Hrubec, M.~Jeitler\cmsAuthorMark{1}, A.~K\"{o}nig, I.~Kr\"{a}tschmer, D.~Liko, T.~Matsushita, I.~Mikulec, D.~Rabady, N.~Rad, B.~Rahbaran, H.~Rohringer, J.~Schieck\cmsAuthorMark{1}, J.~Strauss, W.~Waltenberger, C.-E.~Wulz\cmsAuthorMark{1}
% ...
\vskip\cmsinstskip
% ...
1:~~Also at Vienna University of Technology, Vienna, Austria\\
% ...

\end{sloppypar}
\end{document}